\documentclass[twocolumn,prb,aps]{revtex4}
\usepackage[dvips]{graphicx}
\usepackage{color}
\usepackage{hyperref}
\newcommand{\deps}{\Delta \epsilon}

\begin{document}
\title{Enhancement of the effective disorder potential and the thermopower in Na$_x$CoO$_2$ through the electron-phonon coupling}

\pacs{71.27.+a,71.10.Fd,71.38.-k}

\begin{abstract} 
The effects of an electron-phonon ($e$-ph) interaction on the thermoelectric properties of Na$_x$CoO$_2$ are analyzed.
By means of dynamical mean field theory calculations we find that the $e$-ph coupling acts in a cooperative way with the disorder, enhancing the effective binary disorder potential strength on the Co sites, which stems from the presence or absence of a neighboring Na atom. 
Hence, the inclusion of the $e$-ph coupling allows us to assume smaller values of the binary disorder potential strength  -- which for Na$_x$CoO$_2$ are in fact also more reasonable. 
More generally, we can conclude that the interplay between disorder effects and coupling to the lattice can be exploited to engineer more efficient thermoelectric materials.
\end{abstract}

\author{G.~Sangiovanni$^{(1, \,2)}$, P.~Wissgott$^{(1)}$, F.~Assaad$^{(2)}$, A.~Toschi$^{(1)}$ and K.~Held$^{(1)}$}
\affiliation{
${}^1$Institut f\"ur Festk\"orperphysik, Technische Universit\"at Wien, Vienna, Austria \\
${}^2$Institut f\"ur Theoretische Physik und Astrophysik, Universit\"at W\"urzburg, Am Hubland, D-97074 W\"urzburg, Germany \\
}

\maketitle

Materials with a high thermopower \cite{Mahan} attract strong interest because of their potential use in heat-electricity conversion.
Recently, experimental and theoretical investigations evidenced that materials with strongly correlated electrons  \cite{Paschen} can have very high values of the thermoelectric figures of merit $ZT$, i.e., a high thermoelectric efficiency. 
In this respect,  Na$_x$CoO$_2$ is one of the most promising materials \cite{HighZTNaCoO2} besides presently used semiconductors such as PbTe \cite{PbTeDOS}.
It has been shown  that several conditions are important for getting the high thermopower: a ``pudding mold''-like shape of the band structure \cite{Kuroki07}, strong correlation and also disorder \cite{marianettiPRL98,wissgottPRB82,longpaper,tomczakPRB82}.

Phonons have also been argued to have a positive effect on the thermopower \cite{hochbaumNature451,boukaiNature451}: ``nano phonon engineering'' as in heterostructures, thin films, cage structures and nanowires can indeed overcome the obstacle that a high electrical conductivity is usually connected with a high thermal conductivity, which is inversely proportional to the figure of merit.
However, in materials characterized by the presence of several concurrent physical mechanisms, not only the individual effects, but also the interplay and the possible synergies between them have to be carefully analyzed. 
For instance the cooperation between the electron-phonon interaction and the disorder is in this respect highly unexplored \cite{note_inter}.

The importance of the Na disorder potential was first brought up by Singh and Kasinathan \cite{singhPRL97} who attributed the differences between their local density approximation (LDA) and photoemission experiments \cite{expNaCoO} to the disorder.
The effect of two inequivalent sites on the electronic correlations was then analyzed by Marianetti and Kotliar \cite{marianettiPRL98,furtherDMFT} by means of LDA+dynamical mean field theory (DMFT) \cite{LDADMFT} with the inclusion of the disorder at level of the coherent potential approximation (CPA). 
Further LDA+DMFT studies which focused on reproducing the thermopower of Na$_x$CoO$_2$ \cite{wissgottPRB82,longpaper} analyzed the mechanism behind the influence of disorder on electronic correlations: because of the disorder the lattice sites have locally different occupations and since a fraction of them moves towards half-filling, correlation effects get effectively enhanced.
This increase in correlation leads to stronger electron-hole asymmetry in the LDA+DMFT spectra w.r.t. the LDA and this favors a large thermopower. 

An important point of the LDA+DMFT studies discussed above is that the value used for the disorder potential $\Delta \epsilon \! = \! 0.55$eV is about ten times larger than the one which can be estimated from the energy distance of the LDA bands attributed to the inequivalent sites \cite{singhPRL97}. 
It is therefore evident that electronic correlation and disorder effects are not enough to explain the experimental findings. This leaves the question of which are the relevant interactions to get high values for the figure of merit answered. 

Here we propose the cooperative interplay between electron-phonon interaction and disorder as a new mechanism for increasing the thermopower.
Our findings therefore suggest a very simple route to high figures of merit in strongly correlated materials with a favorable shape of the band structure: Introduce a small amount of disorder in those compounds in which the electron-phonon interaction is also important.
To support our claim we add an electron-phonon coupling to the LDA+DMFT calculations for Na$_x$CoO$_2$, i.e. to the simplest thermoelectric materials with electronic correlation, disorder and favorable shape of the bands around the Fermi level.
We focus on the characteristics of the self-energy that in Refs. \onlinecite{wissgottPRB82,longpaper} were identified to enhance the thermopower.
The goal is to obtain the same features in the presence of an electron-phonon coupling, without the need of assuming unrealistically large values of the disorder strength. 
We emphasize that the resulting phonon enhancement of the thermopower which is introduced here is an additional effect besides the suppression of the thermal conductivity discussed before \cite{hochbaumNature451,boukaiNature451}.

Before we present our results for the specific case of Na$_x$CoO$_2$, let us discuss why we expect in general $e$-ph interaction and disorder to pull together in the same direction: 
A local $e$-ph coupling to the electronic charge gives -- among other effects -- a shift of the local energy level, i.e. an effective shift of the local chemical potential.
The role of such a shift has been investigated by one of us before \cite{hhdoping} in a more general context of DMFT model calculations: For small values of the phonon frequency it is well described by the following linear behavior:
\begin{equation}
\label{mulam}
\mu(\lambda) \simeq \mu - n \lambda W 
\end{equation}
where $n$ is the density, $W$ the electronic bandwidth  and $\lambda$ the dimensionless electron-phonon coupling (see below). 
This means that the $e$-ph coupling $\lambda$ reduces the local chemical potential $\mu(\lambda)$ of the occupied sites ($n>0$).
In the CPA, i.e. the framework used here to treat binary disorder in DMFT, a local potential $\Delta \epsilon$ is added to the sites where disorder is present. 
Sites with and without disorder have hence different occupations. 
Therefore the $e$-ph coupling will act quite differently on them and will ``amplify'' the difference in the level renormalization between the two sites. This simple mechanism results in an enhancement of the disorder effects, as we will argue and also demonstrate numerically below.

We employ a single $a_{1g}$ band model for Na$_x$CoO$_2$ derived from the LDA, supplemented by a local Coulomb interaction $U=3.5\,$eV:
\begin{equation}
H_{e}=-\sum_{ ij,\sigma}t_{ij}c^{+}_{j\sigma} c_{i\sigma}^{\phantom{+}} + U\sum_{ i} n_{i\uparrow}n_{i\downarrow}+\deps\sum_{i\in \text{Vac}} n_i
\end{equation}
Here, $c_{i\sigma}^+$ and $c_{i\sigma}$ create and annihilate an electron on site $i$ with spin $\sigma$, respectively, $n_{i\sigma}:=c_{i\sigma}^+c_{i\sigma}$ is the electron density on site $i$ for the spin $\sigma$ and $n_i$ denotes the electron number operator on site $i$ summed over spin.
As already introduced in Ref. \onlinecite{longpaper}, sites with an adjacent Na atom ($i \in \text{Na}$) are not altered, while a local disorder potential $\Delta \epsilon$ is added to the ``vacuum'' sites ($i \in \text{Vac}$).   

The coupling to the A$_{1g}$  phonon is described by the following Holstein form:
\begin{equation}
  H_{e-ph} =  - g \sum_{i,\sigma}  \left( a_{i}^{+}  +  a_{i} \right) n_{i \sigma}  + \omega_0 \sum_{i}a_{i}^{+} a_{i}^{\phantom +}.
\end{equation}
The electron-phonon coupling is given by $\lambda \! =\! E_{bipol}/W $ with $E_{bipol}\! =\! 2 g^{2}/{\omega_0}$ being the bipolaronic binding energy.  
This Hamiltonian captures the local part of the electron-phonon interaction in Na$_x$CoO$_2$, which as shown in Ref. \onlinecite{donkov}, is the one with the largest coupling. 
The additional physics associated to the non-local part of the electron-phonon coupling lies beyond the scope of the DMFT approach which we will adopt here.

The impurity solver used for the DMFT equations is a continuous-time quantum Monte Carlo in the ``CT-INT'' implementation which can take into account the phonon degrees of freedom.  This algorithm is very well suited to include a local Holstein electron-phonon coupling to an Einstein mode.  
The idea is to integrate out the  bosonic  mode so as to obtain a retarded  density-density interaction.    
This retarded  interaction can readily be taken into account within the framework of  the  CT-INT   algorithm.   
Details  of the  implementation can be found in Ref. \onlinecite{Assaad07} and selected applications in Refs. \onlinecite{assaadPRB78,hohenadlerPRB83}.
The CPA used for the disorder boils down to an arithmetic average of the local Green function weighted by the corresponding Na and vacuum densities $n_{\text{Na}}$ and $n_{\text{Vac}}$.  
For a given value of the disorder potential we adjust the chemical potential $\mu$ such that the total density $n_{\text{tot}}=x n_{\text{Na}} + (1-x)  n_{\text{Vac}}$ equals $1.7$. 

\begin{figure}[th]
\begin{center}
\includegraphics[width=8cm]{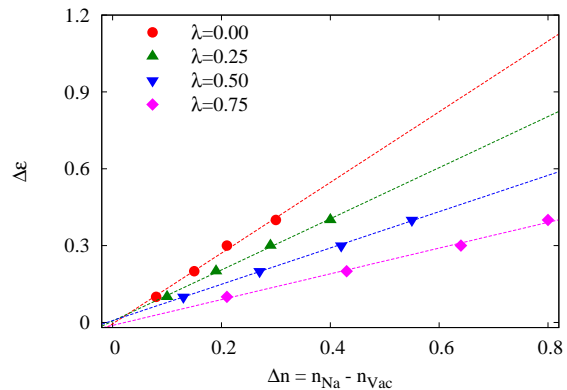}
\caption{$\Delta \epsilon$ as a function of $\Delta n$ for different values of $\lambda$.}
\label{eps_vs_n}
\end{center}
\end{figure}

In the absence of $e$-ph coupling  $\deps$ leads to a local charge density disproportionation $\Delta n \! =\! n_{\text{Na}} - n_{\text{Vac}}$ which, in the considered parameter range, turns out to be linear: 
$\Delta \epsilon = b \Delta n $. 
This is shown in Fig.\ \ref{eps_vs_n} (red circles).
   Using Eq. \ref{mulam}  and for a fixed charge imbalance, we can deduce within the framework of
 the CPA  that  $\Delta \epsilon(\lambda) \!  =\! \Delta \epsilon(\lambda\!=\!0) - \Delta n  \lambda W$, thus leading to 
a $\lambda $ dependence of the coefficient $b(\lambda)  = b  - \lambda W  $.   Hence, in the presence of the 
$e$-ph interaction we conjecture that  
\begin{equation}
\label{deltanl}
\Delta \epsilon =  \left( b - \lambda W \right) \Delta n 
\end{equation}

In other words, for the same given disorder potential $\Delta \epsilon$, a larger charge disproportionation $\Delta n$ is obtained if the $e$-ph coupling $\lambda$ is included. 
This explains the change in the slope upon increasing $\lambda$ observed in Fig.\ \ref{eps_vs_n}, where the results of our calculation are shown.
As we expect from Eq. (\ref{deltanl}) the behavior is linear with and without $e$-ph coupling:
As it should be, all linear fits (dashed lines in Fig. \ref{eps_vs_n}) to the data have an intercept at  $(0,0)$ and the linear form accounts well for the 
numerical data.  
The comparison with Eq. \ref{deltanl} is therefore qualitatively very good, albeit the actual values obtained for the slopes are larger than $b-\lambda W$. 
Even though  Eq. \ref{mulam}  gives a quantitative estimate of the chemical potential,  the compressibility,  which  corresponds to the inverse slope with respect to $n$, comes out less accurately.  
The interplay between electronic and lattice interactions indeed may renormalize an effective electron phonon coupling, going beyond the approximation in  Eq. \ref{mulam}. 

The changes in the slope of $\Delta \epsilon$ $vs$ $\Delta n$ visible in Fig. \ref{eps_vs_n} can be also understood as follows: The compressibility $\kappa \equiv \partial n / \partial \mu$ of a Fermi liquid in the presence of an electron-phonon coupling reads \cite{grilliPRB50,beccaPRB54}:
\begin{equation}  \label{kappa}
\kappa_U(\lambda)  =  \frac{\kappa_U(0)}{1- \kappa_U(0) \lambda W}
\end{equation}
where $\kappa_U(0)$ is the compressibility in the absence of electron-phonon interaction. 
The Hubbard $U$ enhances the compressibility w.r.t. the LDA value (in other words $\kappa_U(0)$ is bigger than the free compressibility given by the value of the density of the states). The $e$-ph interaction then further increases it according to Eq. \ref{kappa}.
A qualitatively similar increase can be observed in the slope of $\Delta n$ as a function of $\Delta \epsilon$ (actually a decrease of the inverse quantity, i.e. of the slope of $\Delta \epsilon$ as a function of  $\Delta n$), as shown in Fig. \ref{eps_vs_n}.
Since this is not the charge response to a uniform change of the chemical potential, we are not dealing with a true compressibility. $\Delta n$ corresponds rather to a ``CPA''-like shift of the relative occupations but the behavior with $\lambda$ is qualitative very similar to that of $\kappa_U(\lambda)$ of Eq. \ref{eps_vs_n}. 
Another difference is also that the true compressibility would not look as linear in $\mu$ as the CPA result for $\Delta n$ as a function of $\Delta \epsilon$.
Concerning the exact values for the slope with and without phonons, let us note that in previous model calculations within DMFT the compressibility was also only qualitatively described by Eq.~\ref{kappa} despite the fact that the chemical potential was very accurately given by the relation Eq.~\ref{mulam} \cite{hhps,hhdoping}.

We have therefore shown that the $e$-ph interaction effects on the effective local potential go in the same direction as those of the disorder. We have understood this in terms of the effect of an $e$-ph coupling on the compressibility of a Fermi liquid.
For the purposes of the present analysis, this observation means that we can expect to obtain similar disorder-induced effects on the self-energy by considering the $e$-ph coupling and smaller (as well as more realistic) values of $\Delta \epsilon$ compared to those used in calculations at $\lambda \! = \! 0$.
Thus, we can conclude that large values of the thermopower can be obtained in the presence of $e$-ph interaction by combining a moderate degree of disorder with electronic correlation.

\begin{figure}[htbp]
\begin{center}
\includegraphics[width=8cm,angle=0]{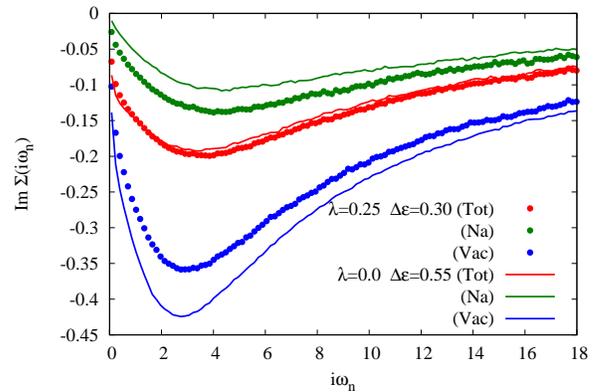}
\caption{Imaginary part of the self-energy on the Matsubara axis for electron-phonon coupling $\lambda\!=\!0.25$ (symbols) and $\lambda\!=\!0.00$ (solid lines).} 
\label{figse1}
\end{center}
\end{figure}

\begin{figure}[htbp]
\begin{center}
\includegraphics[width=8cm,angle=0]{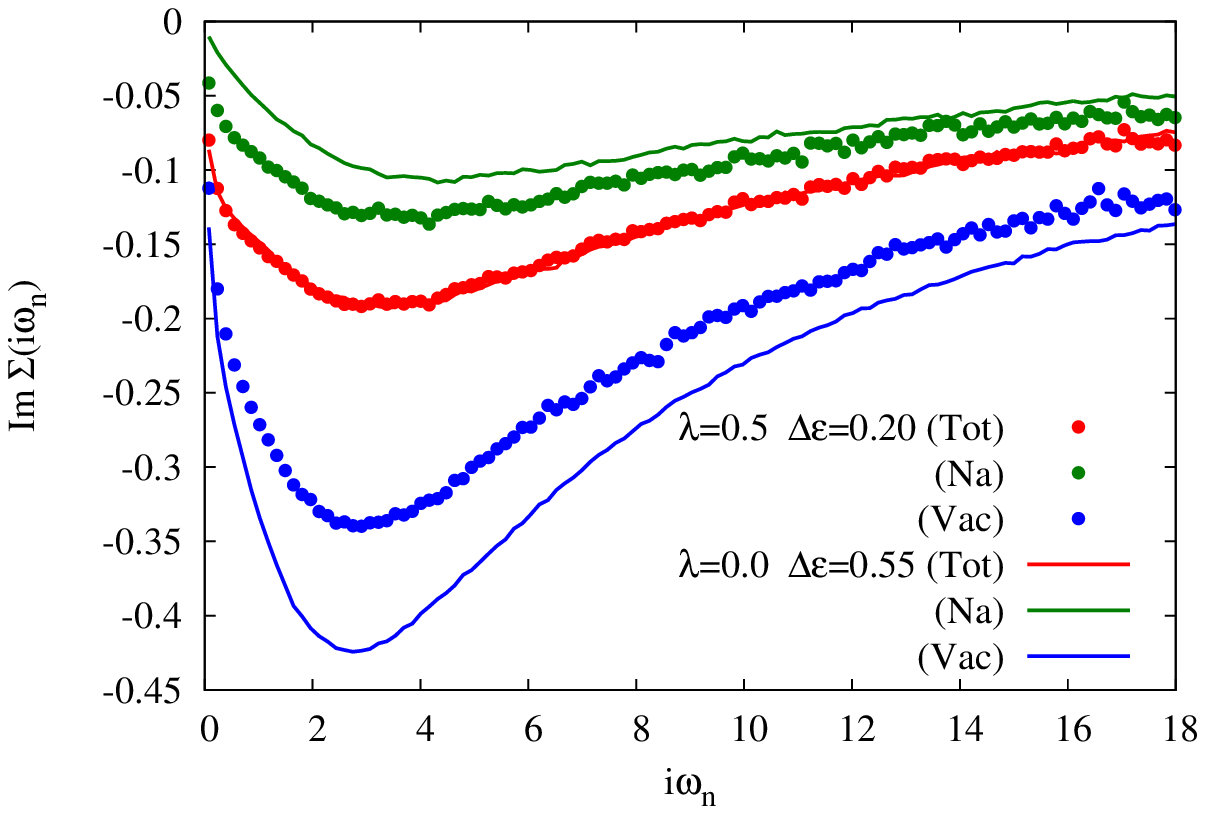}
\caption{Imaginary part of the self-energy on the Matsubara axis for $\lambda\!=\!0.50$ (symbols) and $\lambda\!=\!0.00$ (solid lines).} 
\label{figse2}
\end{center}
\end{figure}

\begin{figure}[htbp]
\begin{center}
\includegraphics[width=8cm,angle=0]{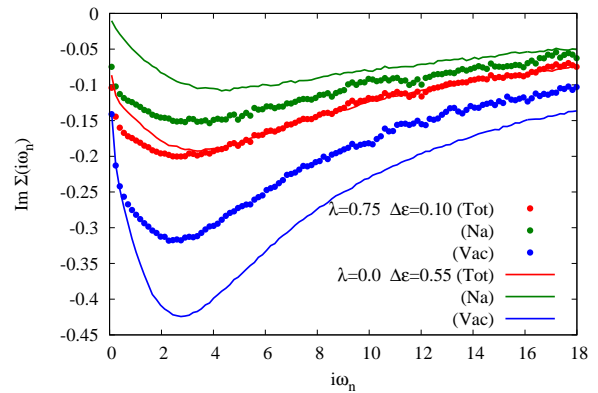}
\caption{Imaginary part of the self-energy on the Matsubara axis for $\lambda\!=\!0.75$ (symbols) and $\lambda\!=\!0.00$ (solid lines).} 
\label{figse3}
\end{center}
\end{figure}

To see this quantitatively we can concentrate on the signatures characterizing the large thermopower in the self-energy of Na$_x$CoO$_2$, namely -- as shown in Refs. \onlinecite{wissgottPRB82,longpaper} --  a non-zero Im $\Sigma(0)$ and a deviation from a linear-dependence of Im $\Sigma(i\omega_n)$ at small $\omega_n$, characterizing Fermi-liquid solutions. 
In Fig. \ref{figse1}, \ref{figse2} and \ref{figse3} we show the self-energy of the whole system (``Tot'') and that of the individual sites with and without Na.
Lines represent the case with no phonons and large strength of the disorder while the symbols indicates the cases with increasing $\lambda$ and correspondingly smaller and smaller $\Delta \epsilon$. 
From Fig. \ref{figse1}, \ref{figse2} and \ref{figse3} it is clear that in the presence of $e$-ph interaction both self-energy features discussed above appear in the ``Tot'' curve for smaller values of $\Delta \epsilon$ than without $e$-ph interaction.
For $\lambda=0.25$, for example, a value of 0.30 instead of 0.55 suffices (Fig. \ref{figse1}) and the effect gets rapidly stronger upon further increasing $\lambda$.

It is also interesting to analyze how the individual self-energies are influenced by the electron phonon interaction: The ``Vac'' self-energy, corresponding to a locally higher density of holes is less strongly modified by phonons at small Matsubara frequencies, the reason for this being that the Hubbard interaction more effectively blocks charge fluctuations at higher densities. The ``Na'' self-energy does get instead more influenced by $\lambda$, reflecting the fact the the ``Na'' sites are locally closer to the low-density hole limit in which the Hubbard $U$ effects vanish. 
In this light, it is possible to also understand the different effectiveness of the $e$-ph coupling for Na$_x$CoO$_2$ and CPA disorder: the weight of the ``Na'' component is large and consequently, a modification of that component, as the one induced by $\lambda$, results in a strong effect for the entire system. 
The picture of two components having locally two different site occupations explains also why the slope of the corresponding self-energies is changed in opposite ways: That of the ``Na'' self-energy increases in absolute value, while that of the ``Vac'' sites decreases. 
In the former the effects of $U$ are indeed smaller and therefore the phonons act in a more conventional way, i.e. they tend to localize electrons (eventually leading to polaron trapping). For the ``Vac'' component, instead, the main effect of the $e$-ph interaction is a reduction of the total repulsion of the system. In other words phonons increase the distance from the metal-insulator transition $U$ values and, as a result, electrons have a smaller effective mass (smaller absolute value of the slope of Im $\Sigma (i\omega_n)$) \cite{hhscreening}.
Both effects can be clearly observed in Fig. \ref{figse1}, \ref{figse2} and \ref{figse3}.

Our analysis of the $e$-ph support to the thermopower of strongly correlated thermoelectric materials with disorder, such as Na$_x$CoO$_2$, would be a mere academic exercise if the $e$-ph coupling would not be strong enough to give rise to a visible effect.
In the literature there are two independent estimates of the $e$-ph coupling in this material.
The first one is a direct LDA calculation within linear-response, where the values of $\lambda$ of 0.21 and 0.29 are reported, for $x\!=\!0.3$ and 0.7 respectively \cite{rueffPRB74}.
The second is an estimate based on electrostatic arguments and a comparison with Raman experiments \cite{donkov}. This, for the (orbital) diagonal part of the $A_{1g}$ oxygen mode gives a value of about 0.044, which summed over all ${\bf q}$ vectors gives about 0.13. 
A factor of two between the two numbers is understandable because first of all the latter is the coupling to just one selected mode and also because it contains a very delicate estimate of the dielectric constant in Na$_x$CoO$_2$. 
In addition, the $\lambda$ estimated in LDA is extracted from the Eliashberg function and contains the phonon frequency renormalization effects, while the $\lambda$ estimated by Donkov, {\it et al.} is the ``bare'' $e$-ph coupling constant (for a discussion about the difference between these types of estimates see Ref. \onlinecite{bauerPRB84}).

If in our scheme we use a value close to the LDA estimate for Na$_x$CoO$_2$, namely $\lambda=0.25$, we get (Fig. \ref{figse1}) strong effects on the self-energy for $\Delta \epsilon \! = \! 0.30$, i.e. almost a factor of two smaller than what used in calculations without $e$-ph interaction.
Our analysis therefore fully supports the claim that there are the conditions to exploit the $e$-ph interaction for enhancing the thermopower of strongly correlated materials, provided that the conditions of a favorable asymmetric shape of the bands is present and a moderate degree of disorder is included.

We would like to thank D. Singh, S. Ciuchi and I. Eremin for useful discussions.
We acknowledge the financial support from the Austrian Science Fund (FWF) through ``Lise-Meitner'' Grant No. M1136 (G.S.), 
F4103-N13 (P.W.), I610-N16 (A.T.) and research unit FOR1346 of the Deutsche Forschungsgemeinschaft and FWF I597-N16 (F.A. and K.H.).


\begin{thebibliography}{99}
\bibitem{Mahan}
G.\ D.\ Mahan,  Solid State Physics {\bf 51}, 81 (1997).
G.\ D.\ Mahan {\em et al.}, Physics Today, March 1997, p.42.


\bibitem{Paschen} S. Paschen, Thermoelectric aspects of strongly
correlated electron systems, in CRC Handbook of Thermoelectrics, Ch. 15,
(ed. D. M. Rowe, CRC Press), Boca Raton, 2005.

\bibitem{HighZTNaCoO2}
M.~Ohtaki, {\em Oxide Thermoelectric Materials for Heat-to-electricity Direct Energy Conversion}, Kyushu Uniersity Global COE Program Novel Carbon Resources Sciences Newsletter, 2010.05.
M.~Ito, T.~Nagira and S.~Hara, Journal of Alloys and Compounds {\bf 408}-{\bf 412}, 1217 (2006). 
K.~Fujita, T.~Mochida and K.~Nakamura, Jpn. J. Appl. Phys. {\bf 40}, 4644 (2001) 

\bibitem{PbTeDOS}
See, e.g., 
J. P. Heremans  {\em et al.}, Science {\bf 321}, 554 (2008).

\bibitem{Kuroki07}
K.\ Kuroki, R.\ Arita, J. Phys. Soc. Jpn. {\bf 76}, 083707 (2007). Also note a related double pudding mold in LiRh$_2$O$_4$, R. Arita {\em et al.} PRB {\bf 78}, 115121 (2008).

\bibitem{marianettiPRL98} C. A. Marianetti and G. Kotliar, Phys. Rev. Lett. {\bf 98}, 176405 (2007).
\bibitem{wissgottPRB82} P. Wissgott, A. Toschi, H. Usui, K. Kuroki, and K. Held, Phys. Rev. B {\bf 82}, 201106(R) (2010).
\bibitem{longpaper} P. Wissgott, A. Toschi, G. Sangiovanni and K. Held, Phys. Rev. B {\bf 84}, 085129 (2011).
\bibitem{tomczakPRB82} J.~Tomczak, K.~Haule, T.~Miyake, A.~Georges and G.~Kotliar, Phys. Rev. B {\bf 82}, 085104 (2010).
\bibitem{hochbaumNature451} A.~Hochbaum, R.~Chen, R.~Delgado, W.~Liang, E.~Garnett, M.~Najarian, A.~Majumdar and P.~Yang, Nature {\bf 451}, 163 (2007).
\bibitem{boukaiNature451} A.~Boukai, Y.~Bunimovich, J.~Tahir-Kheli, J.-K.~Yu, W.~Goddard~III and J.~Heath, Nature {\bf 451}, 168 (2007).


\bibitem{note_inter}{S.~R.~Phillpot, D.~Baeriswyl, A.~R.~Bishop and P.~S.~Lomdahl, Phys. Rev. B {\bf 35}, 7533 (1987), S.~Kumar and P.~Majumdar, Phys. Rev. Lett. {\bf 94}, 136601 (2005), S.~Fratini and S.~Ciuchi, Phys. Rev. B {\bf 72}, 235107 (2005).}

\bibitem{singhPRL97}{D. J. Singh and D. Kasinathan, Phys. Rev. Lett. {\bf 97}, 016404 (2006).}
\bibitem{expNaCoO}H.-B. Yang, {\it et al.} Phys. Rev. Lett. {\bf 95}, 146401 (2005). D.~Qian , {\it et al.} Phys. Rev. Lett. {\bf 96}, 046407 (2006).

\bibitem{furtherDMFT}
Let us also note further LDA+DMFT studies \cite{Ishida05,Liebsch08,Lechermann09} of the one-particle spectrum  without disorder, which might be a more appropriate  description at low temperatures where there is experimental indication \cite{Balsys97,Zandbergen04} of an ordering of the Na atoms.

\bibitem{Ishida05}
H. Ishida \emph{et al}, Phys. Rev. Lett. {\bf 94}, 196401 (2005).

\bibitem{Liebsch08}
A. Liebsch, H. Ishida, Eur. Phys. J. B {\bf 61}, 405-411 (2008).

\bibitem{Lechermann09}
F. Lechermann, Phys. Rev. Lett. {\bf 102}, 046403 (2009)


\bibitem{Balsys97}
R.~J.~Balsys and R.~L.~Davis, Solid State Ionics {\bf 93}, 279-282 (1997).


\bibitem{Zandbergen04}
H.~W.~Zandbergen, \emph{et al.}, Phys. Rev. B {\bf 70}, 024101 (2004).

\bibitem{LDADMFT}
V. I. Anisimov, A. I. Poteryaev, M. A. Korotin, A. O. Anokhin, and G. Kotliar, J. Phys.: Condens. Matter {\bf 9}, 7359 (1997).
A. I. Lichtenstein and M. I. Katsnelson, Phys. Rev. B {\bf 57}, 6884 (1998).
K. Held, Adv. Phys. {\bf 56}, 829 (2007).

\bibitem{hhdoping}G. Sangiovanni, M. Capone and C. Castellani, Phys. Rev. B {\bf 73}, 165123 (2006).
\bibitem{donkov}A. Donkov, {\it et al.}, Phys. Rev. B {\bf 77}, 100504(R) (2008).
\bibitem{Assaad07} F.~F. Assaad and T.~C. Lang, Phys. Rev. B {\bf 76},  035116  (2007).
\bibitem{assaadPRB78} F. Assaad, Phys. Rev. B {\bf 78}, 155124 (2008).
\bibitem{hohenadlerPRB83} M. Hohenadler, H. Fehske and F. Assaad, Phys. Rev. B {\bf 83}, 115105 (2011).
\bibitem{grilliPRB50} M. Grilli and C. Castellani, Phys. Rev. B {\bf 50}, 16880 (1994).
\bibitem{beccaPRB54} F. Becca, M. Tarquini, M. Grilli, and C. Di Castro, Phys. Rev. B {\bf 54}, 12443 (1996).
\bibitem{hhps}M. Capone, G. Sangiovanni, C. Castellani, C. Di~Castro and M.~Grilli, Phys. Rev. Lett. {\bf 92}, 106401 (2004).
\bibitem{hhscreening} G. Sangiovanni, M. Capone, C. Castellani and M.~Grilli, Phys. Rev. Lett. {\bf 94}, 026401 (2005).
\bibitem{rueffPRB74}J.-P.~Rueff, M.~Calandra, M.~d'Astuto, Ph.~Leininger, A.~Shukla, A.~Bosak, M.~Krisch, H. Ishii, Y.~Cai, P.~Badica, T.~Sasaki, K.~Yamada and K.~Togano, Phys. Rev. B {\bf 74}, 020504(R) (2006).
\bibitem{bauerPRB84} J.~Bauer, J.~Han and O.~Gunnarsson, Phys. Rev. B {\bf 84}, 184531 (2011).
\end{thebibliography}
\end{document}